\documentclass[apj]{emulateapj}
\slugcomment{{\sc Accepted to AJ:} November 3, 2007}

\begin{document}

\title{Galaxy counts at 24 microns in the 
SWIRE Fields}

\author{David L. Shupe\altaffilmark{1,2}, 
Michael Rowan-Robinson\altaffilmark{3},
Carol J. Lonsdale\altaffilmark{2,9},
Frank Masci\altaffilmark{1,2}, 
Tracey Evans\altaffilmark{2}, 
Fan Fang\altaffilmark{1,2}, 
Sebastian Oliver\altaffilmark{4}, 
Mattia Vaccari\altaffilmark{3,5}, 
Giulia Rodighiero\altaffilmark{5}, 
Deborah Padgett\altaffilmark{1}, 
Jason A. Surace\altaffilmark{1,2}, 
C. Kevin Xu\altaffilmark{2},
Stefano Berta\altaffilmark{5,9},
Francesca Pozzi\altaffilmark{6},
Alberto Franceschini\altaffilmark{5}, 
Thomas Babbedge\altaffilmark{3},
Eduardo Gonzales-Solares\altaffilmark{7},
Brian D. Siana\altaffilmark{1}, 
Duncan Farrah\altaffilmark{8},
David T. Frayer\altaffilmark{1,2}, 
H.E. Smith\altaffilmark{9}, 
Maria Polletta\altaffilmark{9,10}, 
Frazer Owen\altaffilmark{11},
Ismael P\'erez-Fournon\altaffilmark{12}}

\altaffiltext{1}{Spitzer Science Center, California Institute of Technology, 
314-6, Pasadena, CA, 91125, USA.}

\altaffiltext{2}{Infrared Processing and Analysis Center, California Institute 
of Technology, 100-22, Pasadena, CA, 91125, USA.}

\altaffiltext{3}{Astrophysics Group, Blackett Laboratory, 
Imperial College London, Prince Consort Road, 
London,  SW7 2BW, UK. }

\altaffiltext{4}{Astronomy Centre, Department of Physics \& Astronomy,
 University of Sussex, Brighton, BN1 9QH, UK.}

\altaffiltext{5}{Dipartimento di Astronomia, Universita di Padova,
Vicolo Osservatorio 5, I-35122 Padua, Italy.}

\altaffiltext{6}{Istituto Nazionale di Astrofisica, Osservatorio Astronomico 
di Bologna, via Ranzani 1, I-40127 Bologna, Italy.}

\altaffiltext{7}{Institute of Astronomy, Madingley Road, Cambridge,
CB3 0HA, UK.}

\altaffiltext{8}{Department of Astronomy, Cornell 
University, Space Sciences Building, Ithaca, NY 14853, USA.}

\altaffiltext{9}{Center for Astrophysics and Space Sciences, University
of California, San Diego, La Jolla, CA 92093-0424, USA.}

\altaffiltext{10} {Institut d'Astrophysique de Paris, 98bis, bd Arago - 75014 Paris, France}

\altaffiltext{11}{National Radio Astronomy Observatory, Socorro, NM 87801, 
USA.}

\altaffiltext{12}{Instituto Astrofiscia de Canarias, Via Lactea,
38200 La Laguna, S/C de Tenerife, Spain.}

\begin{abstract}
This paper presents galaxy source counts at 24 microns
in the six {\it Spitzer} Wide-field InfraRed Extragalactic (SWIRE)  fields. 
The source counts are compared to 
counts in other fields, and to model predictions that have
been updated since the launch of {\it Spitzer}.  This analysis confirms 
a very steep
rise in the Euclidean-normalized differential number counts between 2 mJy
and 0.3 mJy.  Variations in the counts between fields show the
effects of sample variance in the flux range 0.5-10 mJy, up
to 100\% larger than Poisson errors.
Nonetheless, a ``shoulder'' in the normalized counts persists
at around 3 mJy.
The peak of the normalized counts at 0.3 mJy
is higher and narrower than most models predict.
In the ELAIS N1 field,
the 24 micron data are
combined with {\it Spitzer}-IRAC data and five-band optical
imaging, and these bandmerged data are fit with photometric
redshift templates.  Above 1 mJy the counts are dominated by 
galaxies at $z<0.3$.   By 300 microJy, about 25\% are between
$z\sim$0.3-0.8, and a significant fraction are at
$z\sim$1.3-2.  At low redshifts the counts are dominated by
spirals, and starbursts rise in number density to outnumber the
spirals' contribution to the counts below 1 mJy.
\end{abstract}


\section{Introduction}\label{intro}

Galaxy counts from IRAS first hinted at strong evolution of luminous
infrared galaxies \citep{hacking87, lonsdale93, gregorich95}.
Mid-infrared surveys made with the ISOCAM instrument at 15
microns revealed an excess population at z$\sim$0.8 
\citep{Elbaz99,Elbaz02,franceschini01}.
This population was detected in part by the redshifting of the
strong 7.7 micron PAH feature into the 15 micron band.

The high sensitivity at 24 microns of the MIPS instrument
\citep{rieke04} onboard {\it Spitzer} allows infrared-luminous
galaxies to be traced to even higher redshifts.  Sensitivity
to obscured starbursts at z$\sim$2 is enhanced due to redshifting of the 7.7
micron feature into the 24 micron band.  First source counts
have been presented by \cite{Chary04} for the deep
Great Observatory Origins Deep Survey (GOODS) 
test field, by \cite{marleau04} for the First-Look
Survey (FLS) and the  GOODS
 test field, and by \cite{Papovich04}
for a range of Guaranteed Time Observer (GTO) 
fields plus the GOODS test field.  These
counts show a steep bump in the normalized source counts as fluxes
decrease from several mJy to about 300 $\mu$Jy, that is 
attributed to the population discovered by ISOCAM at z$\sim$0.8 plus
a previously unseen population of galaxies
at z$\sim$1-2.  At still fainter fluxes, the normalized counts
drop off to the {\it Spitzer} confusion limit \citep{Rodighiero06, Chary04,
marleau04, Papovich04}.

This paper presents a detailed analysis of 24 micron source counts
from the large fields observed for the {\it Spitzer} Wide-area
InfraRed Extragalactic (SWIRE) program.  The SWIRE survey is
the largest of the extragalactic {\it Spitzer} Legacy Science programs, comprising
mapping of 49 square degrees at wavelength from 3.6 to 160 microns
\citep{lonsdale03,lonsdale04}.  This survey is designed to dramatically
improve our understanding of galaxy evolution, including the
history of star formation, the assembly of stellar mass
in galaxies, the
nature and impact of accretion processes in active nuclei, and
the influence of environment on these processes at angular scales
up to 3 deg (such as the study of clustering of ultraluminous infrared
galaxies by \citet{Farrah06}).  The SWIRE dataset encompasses six fields and so is
an excellent resource for quantifying the effects of sample
variance. 

The goal of this paper is to present accurate
{\it Spitzer} 24$\mu$m counts in the flux ranging from 50 mJy to $\sim$300$\mu$Jy.
A detailed model-based interpretation of the counts is beyond the scope
of this paper.  However, the composition of the counts is explored
for the ELAIS N1 field using 
the IRAC data at 3.6-8$\mu$m \citep{Surace06} as well
as high-quality optical imaging.  
This merged dataset provides
 insights into galaxy SEDs and populations \citep{mrr05, Polletta07},
 and is used in this paper
to break down the 24 $\mu$m source counts by population and redshift.

The paper is organized as follows: The SWIRE observations and
data processing are described in Section \ref{sec:data}.
The derivation and validation of source counts
are presented in Section \ref{sec:counts}.   Comparison with models, 
including recent models updated
using {\it Spitzer} counts, is in Section \ref{sec:models}.  For the
ELAIS N1 field, optical
identifications, photometric redshifts, and template-fitting
lead to source counts subdivided by redshift and galaxy type
in Section \ref{sec:composition}.

\section{SWIRE 24 micron observations and data processing}
\label{sec:data}

\subsection{Observations}
\label{sec:observations}
The {\it Spitzer} mapping observations of the six SWIRE fields were carried
out between December 2003 and December 2004.  For the mapping at
24, 70 and 160 $\mu$m, the following observing strategy using the MIPS
instrument \citep{rieke04}  was implemented.
The MIPS Scan Map Astronomical Observing Template
 was used with medium scan rate.  The spacing
between scan legs was 148 arcseconds, approximately half of the width
of the 5\arcmin$\times$5\arcmin\ field of view (FOV) of the 24$\mu$m detector array.  
Three-degree-long scan legs were used for most Astronomical
Observing Requests (AORs). Five scan legs were typically used per AOR,
resulting in a cross-scan size of about 25\arcmin.  The AORs were arranged into two
coverage sets, separated by half of an FOV in the cross-scan direction.  
 Adjacent AORs within a coverage set were
overlapped sufficiently to ensure no gaps in coverage of the 70$\mu$m array.
The approximate dimensions of the maps are given in Table \ref{tab:photounc}.
\begin{table}
\caption{Area and photometric uncertainty for each field}
\label{tab:photounc}
\begin{tabular}{lcll}
\\
\hline
Field & Size & Area & $\sigma$ \\
 ~ & $[deg\times deg]$ & $[deg^2]$ & (microJy) \\
\hline
ELAIS N1 & $3.1\deg \times 3.0\deg$ & 9.64 & 38.8 \\
ELAIS N2 & $1.9\deg \times 2.5\deg$ & 4.87 &  37.6 \\
Lockman  & $3.7\deg \times 3.0\deg$ & 11.03 & 41.8 \\
CDFS &        $2.6\deg \times 3.0\deg$ & 7.67 & 38.5 \\
XMM-LSS & $3.1\deg \times 3.0 \deg$ & 8.34 & 48.2 \\
ELAIS S1 & $2.3\deg \times 3.0\deg$ & 7.03 & 51.2 \\
\hline\\
\end{tabular}
\end{table}
With this observation layout, the typical coverage per point is
44 Basic Calibrated Data (BCD) images, each with exposure time of
4 seconds, for a total of 160 seconds of integration time
per point at 24$\mu$m, 80 seconds per point at 70$\mu$m, and 
16 seconds per point at 160$\mu$m.  Overlap between rotated scans yielded a higher
coverage in portions of each map.  

Some regions received different coverage levels than described above.
A 0.5 square degree region of the Lockman field was observed as part
of Legacy validation operations, for a total integration time
about 1.5 times the nominal ones described above. In the Lockman and CDFS fields,
the 24 $\mu$m maps include regions of higher coverage around
the Guaranteed Time Observation fields, where scans were initially
embargoed.
The ELAIS S1 field received only
one coverage, and consequently half the nominal observing time per point,
due to observing time constraints.  Part of the ELAIS N1
field, which was missed in January 2005 due to an observatory
standby event, was filled in in July 2005, leading to some
irregularities in field shape and coverage.  We have accounted
for different coverage levels and sensitivities in our analysis.

\subsection{Data processing}
\label{sec:processing}

The 24 micron raw data were processed by the S10.5 version
of the SSC pipelines.  The flat-field images used were
made separately for each scan mirror position, using SWIRE
scan map data.  The SWIRE processing began with the BCD (Basic
Calibrated Data) images.  For all fields, to even out variations in the
background, the median of all the pixels in each image was subtracted from 
each pixel in that image.
In the XMM-LSS field, removal of dark-latent
artifacts was necessary, owing to the proximity of the
very bright source Mira to the MIPS mapping region.
For each scan leg, a self-calibration
flat image was produced from the median of all uncorrected 
BCDs in that
leg,  and was normalized and divided into those BCDs before
the background-subtraction step.  This technique
was successful because the intensity of these latents did
not change appreciably over the 30-minute duration of each
scan leg.  

To make maps, the corrected BCDs were coadded into large
mosaics, using the SSC's MOPEX software \citep{makovoz}.
The mosaics were all made with 1.2\arcsec\ pixels.
Cosmic rays were removed using multi-frame outlier rejection
with a threshold of 2.5 $\sigma$.

Source extraction was performed on the mosaic
using SExtractor \citep{Bertin96}.   A local 
background of size 128x128 pixels was computed for the maps,
and all the noise calculations were weighted by the inverse
square of the coverage map.  Photometry was output 
as measured in several apertures and by various extended source
techniques.  For the present study, we have used
photometry in a 5.25\arcsec\ radius aperture for point sources, and the Kron
flux \citep{Kron80} for extended sources.  Aperture corrections were computed
from comparing smaller apertures to a 30.6-arcsecond-diameter
aperture.  Then, an additional correction of 1.15 was applied
to both aperture and Kron fluxes, to match the procedure used
by the SSC pipelines.  This additional factor is derived from
the fraction of light outside the 30.6\arcsec\ aperture in
theoretical TinyTim models of the point spread function.
The resulting overall correction for the 5.25\arcsec\ aperture
is 1.78.

The calibration supplied with the data assumes a source spectrum of
a 10,000 K blackbody \citep{mipshandbook}.  Galaxies detected
at 24 $\mu$m generally have an SED slope of $F_\nu \sim \nu^{-1}$ or
redder \citep{mrr05}.   Accordingly, we have color-corrected
our photometry to an assumed source spectrum of $F_\nu \sim \nu^{-1}$
by dividing by a factor of 0.961.  Additionally, we have multiplied
by a factor of 1.018 to account for a small change in the 24 $\mu$m
calibration factor made in December 2005 \citep{mipshandbook}.

For this work, sources are assigned the Kron flux if the SExtractor
ISO\_AREA value is at least 100 pixels, and the stellarity is less than
0.8.  All other sources are considered point sources, and the 
5.25\arcsec\ radius aperture measurement is used.

\subsection{Data quality assessment}
\label{sec:quality}

Estimating photometric uncertainties for extractions from mosaicked
images is complicated by correlations between pixels.  To estimate
the uncertainty in each field, we measured fluxes
in thousands of randomly placed apertures that fell in regions
of typical coverage.  Gaussians were fit to
the distribution of fluxes in these apertures, yielding robust
uncertainty estimates.  The photometric uncertainties in the 5.25\arcsec\ 
aperture are given in Table \ref{tab:photounc}. The areas in each
field used in this work are also given.
The relatively high noise value for the XMM-LSS field is due to its
low ecliptic latitude, leading to a high zodiacal background
relative to the other SWIRE fields.  The noise
in ELAIS S1 is higher because this field received half the integration time
of the other fields.

\begin{figure*}
\epsscale{0.8}
\plotone{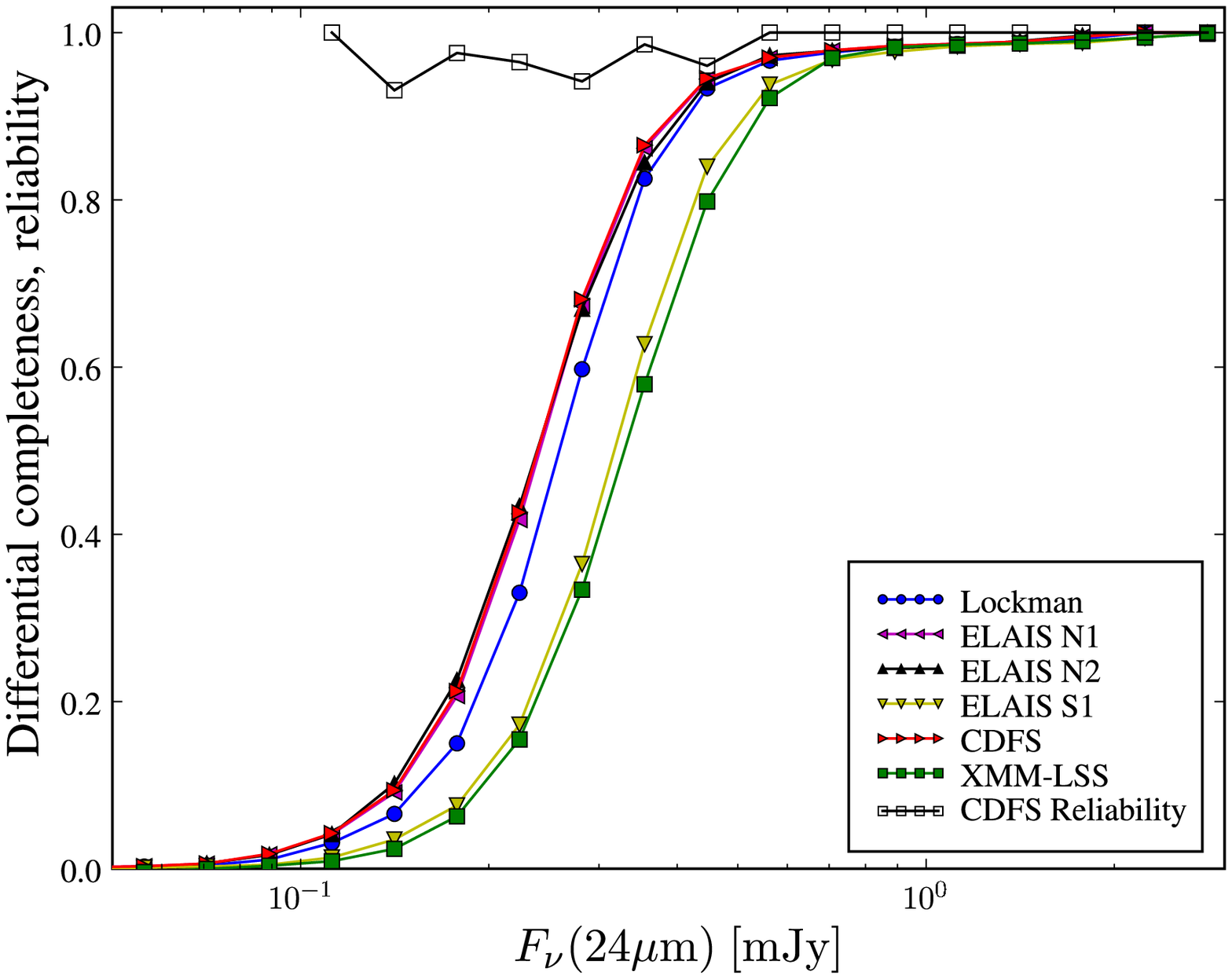}
\caption{{\it Filled symbols:} Differential completeness from simulations
for ELAIS N1, ELAIS N2 and XMM-LSS fields.  The curves for the other three fields were
obtained by shifting the EN1 curve as described in the text. {\it Open squares:} Differential reliability defined
as the fraction of SWIRE sources found in the deeper GTO data in CDFS.}
\label{fig:completeness}
\end{figure*}

We have estimated completeness by inserting simulated sources into
the ELAIS N1, ELAIS N2 and XMM-LSS fields, and computing the fraction 
that are recovered.  The completeness curve shifts to
higher fluxes for  XMM-LSS, approximately following the ratio
of noise levels in Table \ref{tab:photounc}. We have accordingly
shifted the ELAIS N1 curve by the ratio of noise levels to estimate
completeness in Lockman, ELAIS S1 and CDFS (Figure \ref{fig:completeness}).
In the best fields, 70\% completeness is attained
for flux densities above 0.35 mJy.  Based on the noise estimates
in Table \ref{tab:photounc}, this flux density corresponds to a
signal-to-noise ratio of about 7.   Ordinarily the completeness would
be expected to be higher at this SNR, but in our case the SExtractor
detection parameters used have limited the completeness more than
the noise has.

A direct estimate of completeness may be made by
comparing the SWIRE data with deeper data.  We have made such a comparison
by processing GTO data in the CDFS field using the same methods used for the SWIRE data, 
and matching the extractions with a 3.6\arcsec\ 
search radius.  The completeness obtained from this comparison is in good agreement with
the simulation results.

The comparison to deeper data is also useful as an assessment of reliability.
The ratio of all matched sources to all extracted SWIRE sources is also
plotted in Figure \ref{fig:completeness}.  The reliability defined in this
way is better than 98\% in all bins down to 0.35 mJy.

%
%

\section{SWIRE number counts}
\label{sec:counts}

\begin{figure}
\epsscale{1.3}
\plotone{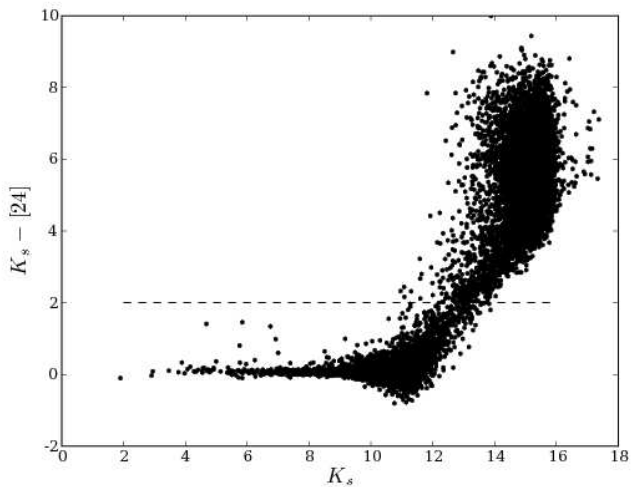}
\caption{Color-magnitude diagram of 2MASS $K_s$ and $[24]$ for all
six SWIRE fields.  The dashed line shows the color cut used to separate
out the stellar contribution (below the line) to the 24$\mu$m number counts.}
\label{fig:k24colormag}
\end{figure}

Although the contribution of stars to the 24 micron number counts
is small at faint (sub-mJy) fluxes \citep{marleau04, Papovich04}, 
it must be taken into account at fluxes of several mJy and brighter.  We have used
2MASS data to identify stars.  Our 24$\mu$m extractions were matched
against the 2MASS catalog using a 3\arcsec\ search radius.  The 
resulting $K_s-[24]$ color-magnitude diagram is shown in Figure
\ref{fig:k24colormag}. $[24]$ is the 24$\mu$m magnitude,
where we have used a zeropoint of
7.43 Jy in our color-corrected flux units (corresponding to 7.14 Jy
for a 10,000 K blackbody spectrum, see \citet{mipshandbook}).  
The $3\sigma$ limit for our
most sensitive fields corresponds to [24] = 12.0, well above the
2MASS completeness limit in unconfused sky of $K_s\sim 14.3$
\citep{Skrutskie06}.  
For our high
Galactic latitude fields, the 2MASS survey is sensitive enough to
detect all stars in the SWIRE 24 $\mu$m catalogs.
The dashed line in the figure shows the simple color cut
used to identify stars for the purpose of this statistical study.
A similar color discriminant was used by \citet{Vaccari05} with ISO
15 $\mu$m fluxes.

\begin{figure*}
\plotone{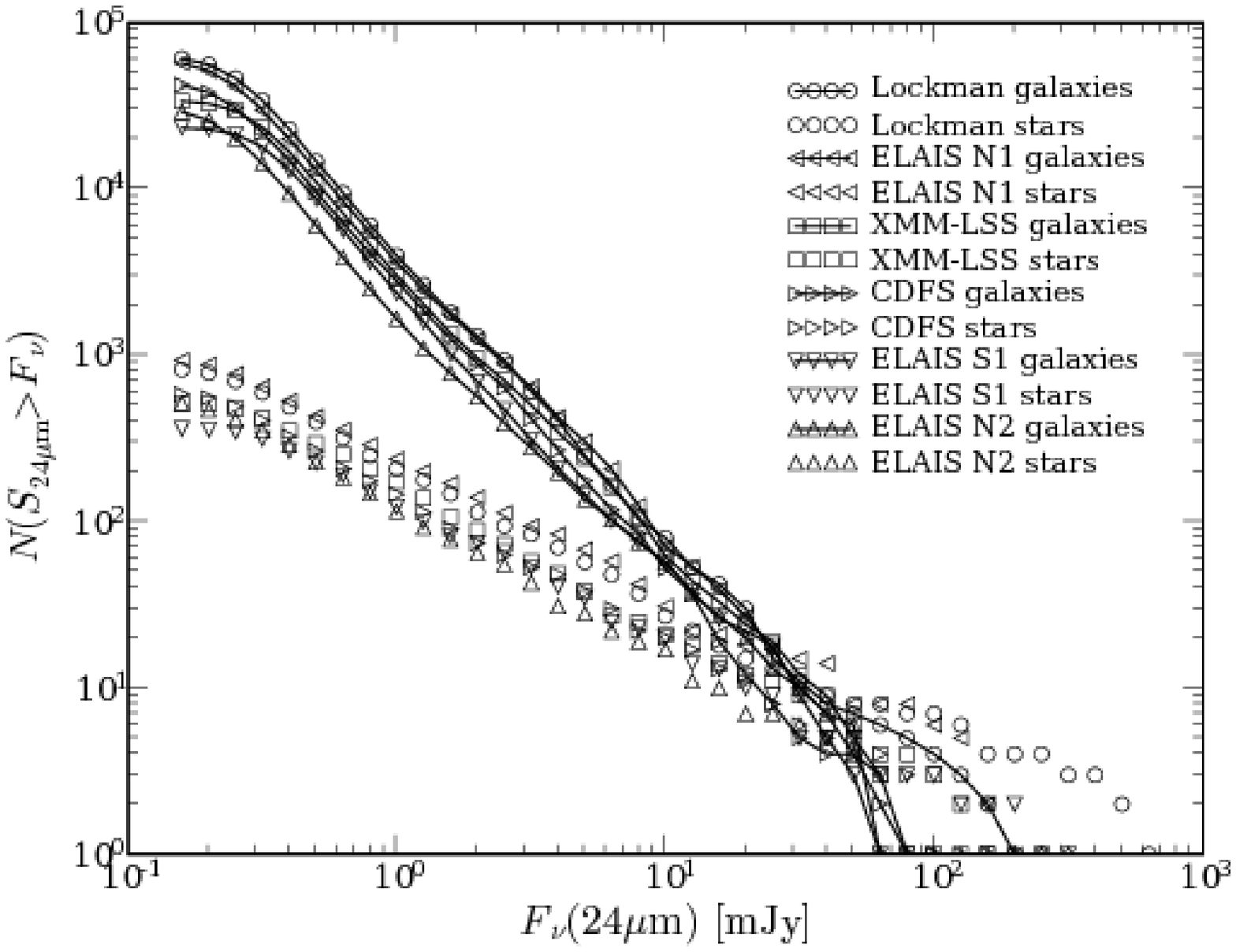}
\caption{Integral counts for the contiguous regions in each SWIRE field.
Galaxy counts are the symbols connected by lines, while stellar counts are 
unconnected.
Note that the counts have not been normalized by area so as to provide
some separation between the distributions.}
\label{fig:integralraw}
\end{figure*}

Integral counts for the stars and galaxies are shown in Figure
\ref{fig:integralraw}.
The integral counts here are the total in each field, without normalization
by the areas in Table \ref{tab:photounc}.  The contribution of stars and
galaxies to the number counts is approximately equal at 30 mJy.

\begin{figure*}
\epsscale{1.0}
\plotone{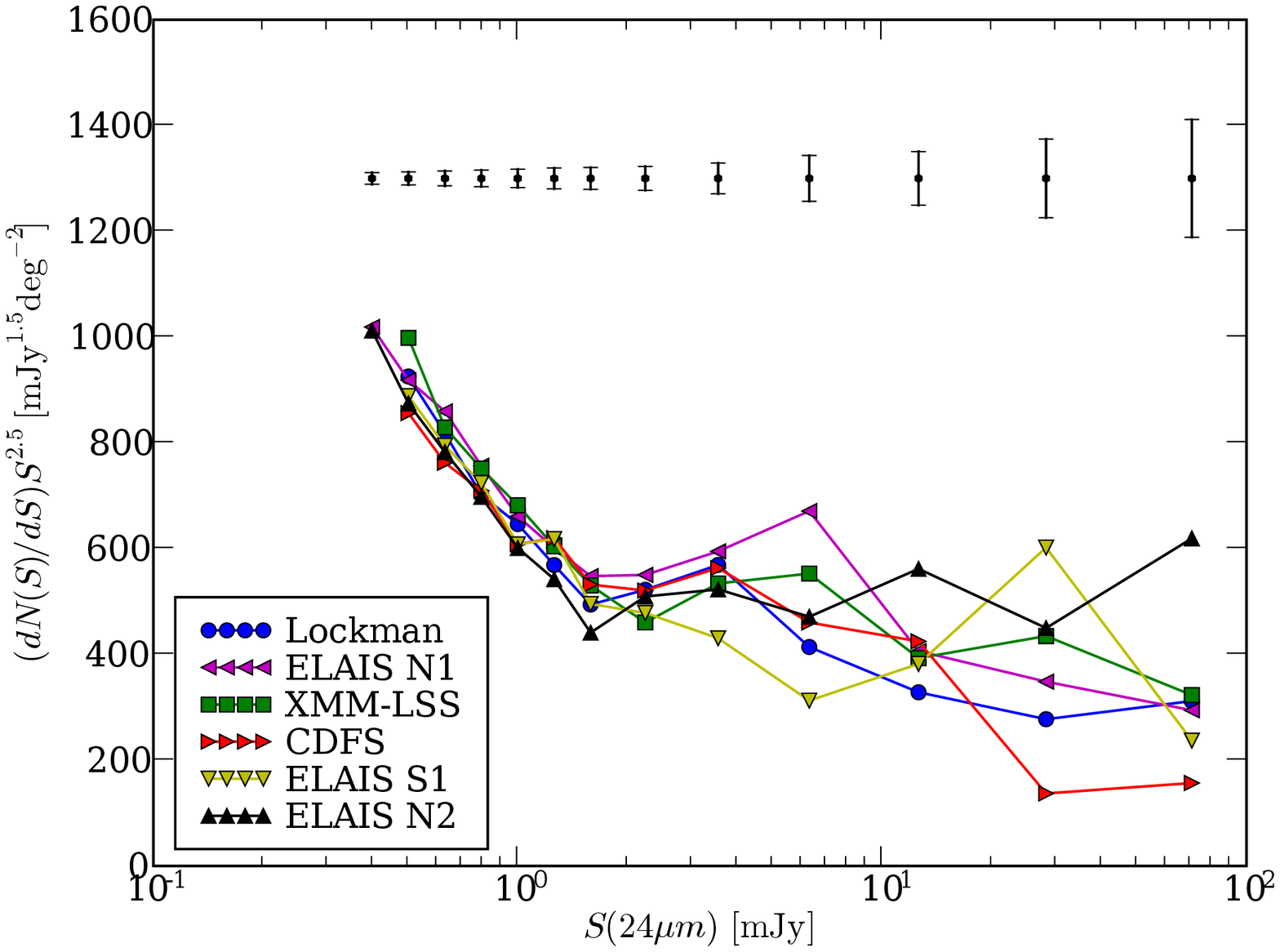}
\caption{Normalized differential number counts for the six SWIRE fields.
For clarity, representative Poisson error bars are shown at the top as
computed for the ELAIS N1 field.}
\label{fig:normswire}
\end{figure*}

Euclidean-normalized number counts were obtained by computing differential counts
with each source individually weighted by a factor of flux raised to the power 2.5.
We have plotted these counts for all fields in Figure \ref{fig:normswire}.
A completeness correction has been applied for each field, using the
curves shown in  Figure \ref{fig:completeness}, shifted for each source
by the square root of the ratio of the coverage to the mean coverage. 
Where the completeness correction becomes large, as in the faintest bin shown
for ELAIS S1, the counts are under-corrected.  

The field-to-field comparison shows differences on the order of 10\%
in the submJy region.  At these fluxes there are of order 10,000 galaxies
per field (Figure \ref{fig:integralraw}) so that counting statistics can
account for only a fraction of this difference.  Even more
striking is the large variations in the vicinity of 3 mJy.  Taken together,
these differences are almost certainly due to sample variance, significant
even on 3-degree scales.

%
%
%
%
%

\begin{figure*}
\plotone{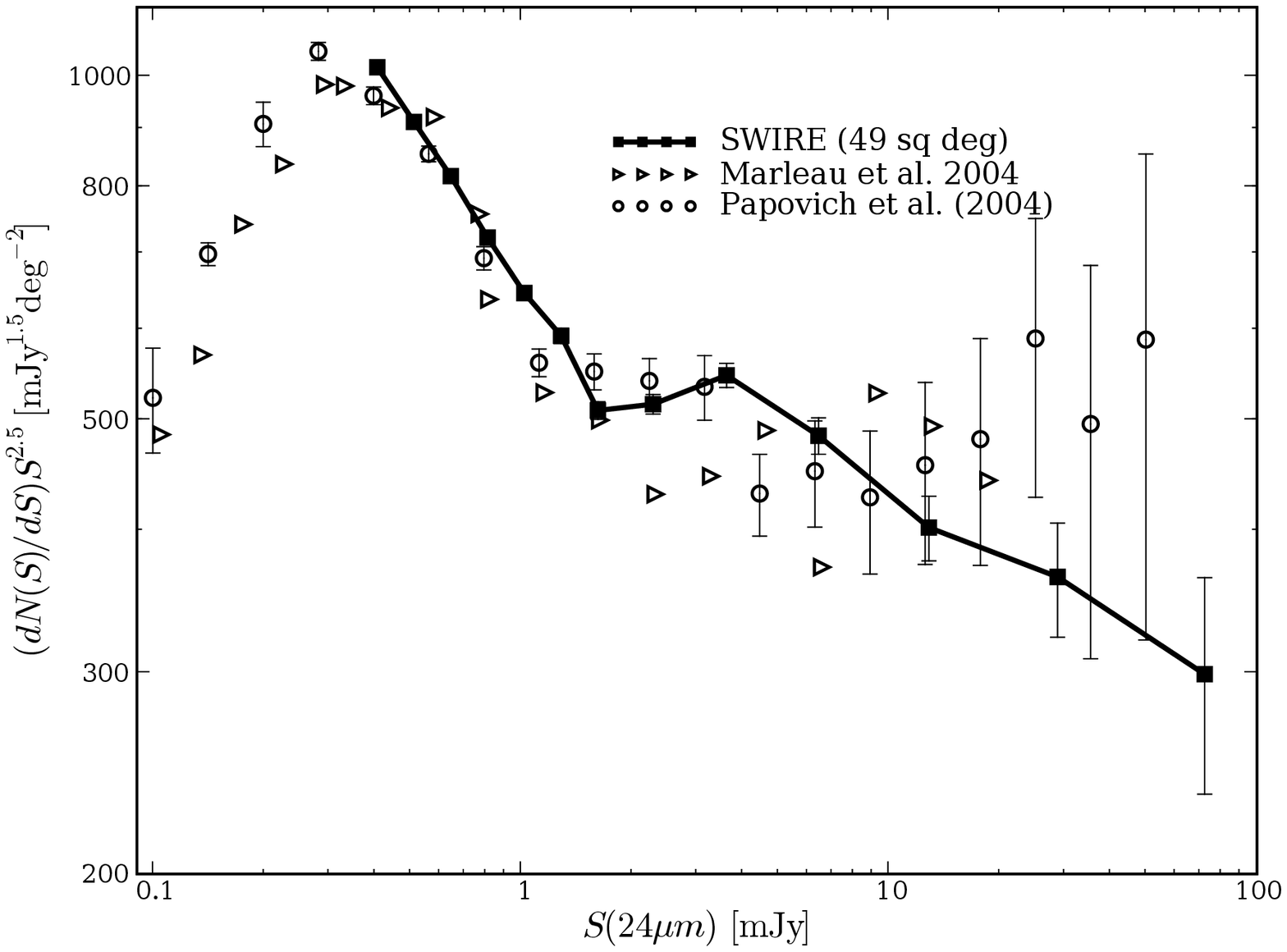}
\caption{Plot of normalized number counts from the average of all
SWIRE fields,
along with those of \citet{Papovich04} and \citet{marleau04}.}  
\label{fig:countsonly}
\end{figure*}

Total SWIRE counts were computed from the combined sample with corrections
for incompleteness applied.  Uncertainties
in the counts
were estimated using a bootstrap technique, in which samples of
identical size to the original are drawn with replacement from the
total sample.
The standard deviation of the counts derived from the bootstrap 
samples provides
the error bars.  The uncertainties are generally quite similar to what
is expected from Poisson statistics, but are somewhat larger at higher
flux levels. Since the counts are derived from six widely spaced fields,
sample variance effects should be averaged out.  The
total SWIRE counts are shown in Figure \ref{fig:countsonly} 
and are tabulated in Table \ref{tab:counts}.
\begin{table}
\caption{Total SWIRE 24 $\mu$m counts}
\label{tab:counts}
\begin{tabular}{rrr}
\\
\hline
Average $F_\nu$ &  Normalized counts & Uncertainty \\
 (mJy)          &  (mJy$^{1.5}$ deg$^{-2}$) & (mJy$^{1.5}$ deg$^{-2}$) \\
\hline
        0.405 &       1018.81 &          4.93 \\
        0.510 &        912.33 &          5.34 \\
        0.642 &        817.73 &          6.15 \\
        0.809 &        722.46 &          6.82 \\
        1.018 &        645.91 &          7.81 \\
        1.282 &        592.51 &          8.97 \\
        1.613 &        509.69 &          9.08 \\
        2.279 &        516.15 &         10.21 \\
        3.612 &        547.15 &         13.38 \\
        6.423 &        484.33 &         17.76 \\
       12.816 &        402.43 &         26.17 \\
       28.691 &        364.24 &         41.77 \\
       72.069 &        299.29 &         64.37 \\
\hline\\
\end{tabular}
\end{table}
We have also plotted the counts 
for the FLS \citep{marleau04} and GTO fields
\citep{Papovich04} for comparison.  To put the Papovich counts on the SWIRE calibration,
we multiplied the fluxes from that work by a factor 1.059 to account for our color
correction, the change in calibration factor \citep{mipshandbook},
 and a very slight difference in aperture corrections.  The FLS counts
were shifted assuming a flux scale fainter than ours by a factor of 1.11, due to
the color correction, the change in calibration factor,
 and our 5\% larger aperture correction for a 30.5\arcsec\ 
diameter aperture.  The steep rise in the normalized
counts found by these works is confirmed by our results.
There are some  differences at brighter
flux levels, but these appear to be within the range of the SWIRE field-to-field
variations shown by the error bars. Also of note is a ``shoulder'' at
about 3 mJy, seen in the
total SWIRE and \citet{Papovich04} counts, though not in the \citet{marleau04}
counts, and also not in the ELAIS S1 counts (Figure \ref{fig:normswire}).

\section {Comparison of counts with models}
\label{sec:models}
\begin{figure*}
\plotone{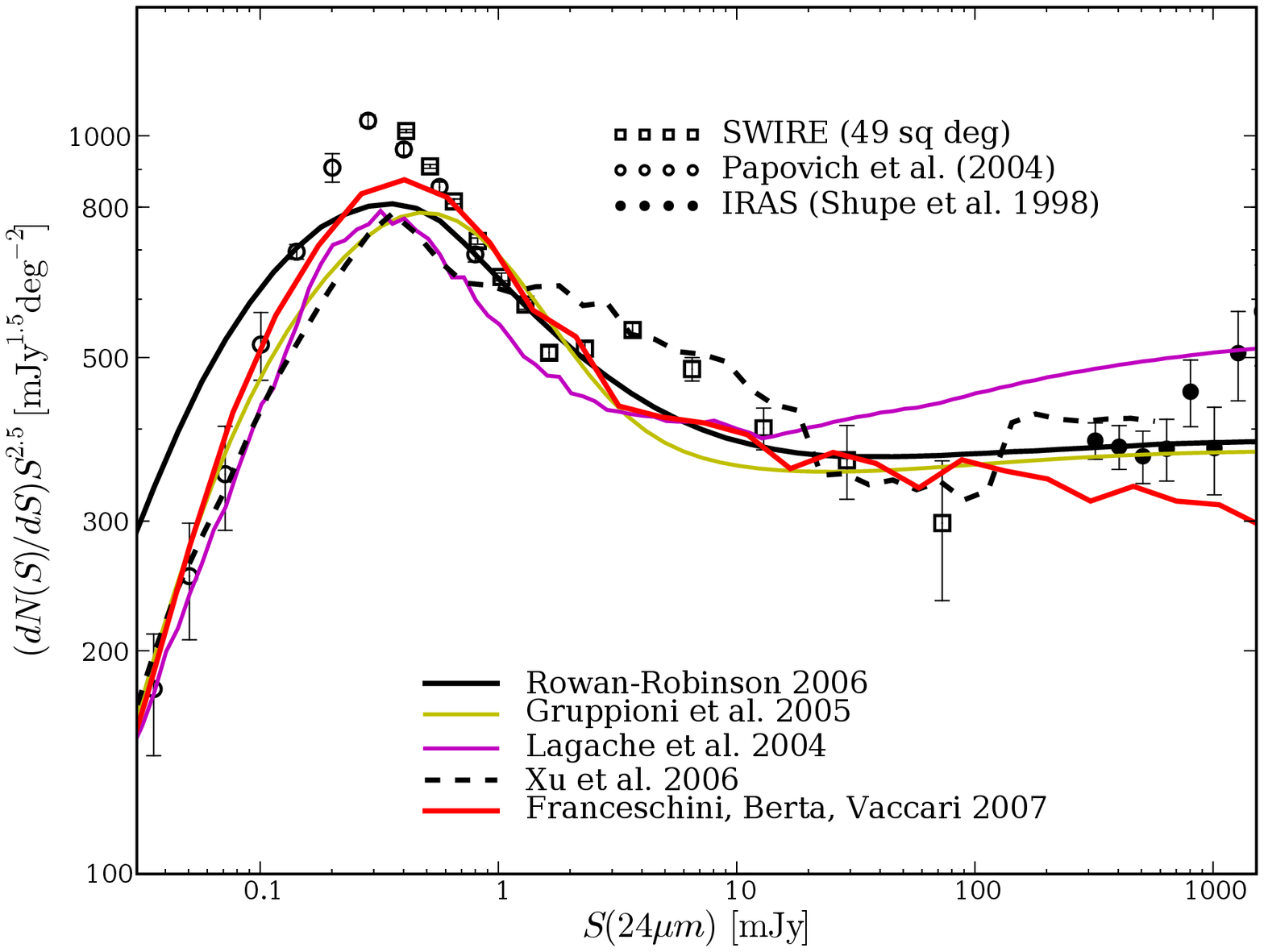}
\caption{Plot of normalized number counts with several models
overlaid.}
\label{fig:counts_models}
\end{figure*}

A plot of the normalized counts with several models overlaid is
in Figure \ref{fig:counts_models}.  The counts used include the
SWIRE counts, the \citet{Papovich04} counts at fainter fluxes, and
IRAS counts from the sample of \citet{Shupe98} corrected to
the mean MIPS wavelength of 23.7 $\mu$m.  The fitting of
several pre-{\it Spitzer}-launch models have already been discussed
in \citet{Chary04} and \citet{Papovich04}.  Here we focus on
more recently available models.  One of these models is described
in detail in \cite{Lagache04} and is not outlined here.  
We give a thumbnail sketch of 
the other models in turn.

\subsection{Details on individual models}

The \citet{Gruppioni05} model is based
on starburst+cirrus galaxies from \citet{Pozzi04} and AGN
components from \citet{Matute06}.  These works
are 15 micron models based on data from the ELAIS survey
\citep{mrr04}.  They
have been converted to 24 microns using the appropriate ratio
of 24$\mu$m to 15$\mu$m of each population.  Though this work was
published after the launch of {\it Spitzer}, it did not involve an {\it a posteriori}
correction.


The predicted 24$\mu$m counts by \citet{franceschini07} are
based on an IR multi-wavelength evolutionary model fitting
a variety of data from Spitzer, ISO, SCUBA, and reproducing
data on the diffuse far-IR background observed by COBE. The model
basically adopts, in addition to normal spiral galaxies and
type-1 quasars, two populations of starbursting galaxies
with different luminosity functions and evolution rates.
One population consists of moderate-luminosity starbursts with
peak activity at z$\sim$1 (the LIRG population), the second contains
more luminous objects (the ULIRG and HYLIRG) with maximum
activity phase at z$\sim$2. This evolutionary bimodality is
apparently required for a combined fit of the multi-wavelength
counts and redshift distributions.

The \citet{Xu06} model includes a new evolution model for dusty galaxies
and the E2 model for the E/S0s evolution \citep{Xu03}..
The new dusty galaxy evolution model is a modified version 
of model S1, the starburst-dominant model, in \citet{Xu03}.
The definitions of galaxy populations (starbursts, normal-disks and AGNs) 
are the same as in model S1, as are the evolution parameters of
the normal-disks and AGNs. Both the luminosity evolution rate
and density evolution rate of the starburst galaxies are on
the order of $(1+z)^3$ for $z\la 1$ and $(1+z)^{-1.5}$
for larger redshifts. The major changes are in the SEDs of the starbursts
and normal-disks: For both populations
the strength of the PAH features in the wavelength range of
$5\mu m < \lambda < 12\mu m$ is raised by a factor of 2.

The model of \citet{mrr07} is a modification 
of the counts model of \citet{mrr01}.  It retains the four infrared
SED types of the latter, with no modification of the SEDs
(unlike the new models of \citet{Lagache04} and \citet{Xu06}).  The four
types still all undergo strong luminosity evolution but the evolution is
allowed to be at a different rate in the different components.  The 
evolution of the cirrus (quiescent) component is now  shallower than 
the evolution of  the M82 starburst component, and is considerably 
steeper for the Arp 220 (high optical depth starburst) component.  The 
evolution function is also modified so that the evolution is less steep 
at z $<$ 0.5 (as in the \citet{Lagache04} model).  The stronger evolution 
in the Arp 220 component makes the model similar to those of 
\citet{franceschini97}, \citet{Xu98} and \citet{dole03}, but it still has the 
distinct feature of strong evolution even in the quiescent component. 

\subsection{Comments on model fits}

It is difficult to fit a steep bump in the normalized counts
without restricting the luminosities in the models to a narrow
range.  That said, the models are successful in peaking at about
the same flux level (0.3 mJy) as the source counts.  The counts
rise rather more steeply with decreasing flux, however, so that
they peak with counts about 25\% higher than the models predict.
Some of this difference is due to the color correction and calibration
changes applied to the counts, which increase flux by about 5.9\%
for the \citep{Papovich04} counts, and increase normalized counts
by about 13\%.    The model curves could be scaled up by this
factor, but in many cases would overpredict brighter sources.

Models which allow evolution of only the more
luminous sources, including the possibility of discontinuous changes
in the evolution rate, or large changes to source SEDs, have
less  difficulty in fitting the counts.  By contrast, models
which use smoothly-changing evolution rates, 
like those of \cite{mrr01} and \cite{mrr07},
have more
difficulty.  It is important to establish what really is needed
to account for the observed counts.  It may be that PAH emission and
silicate absorption features at $z\sim 2$ \citep{Houck05}, or even
silicate emission from AGN \citep{Hao05}, are stronger or play a
larger role than is assumed by the models. 

\section{Composition of 24$\mu$m counts in ELAIS N1}
\label{sec:composition}

We now turn to an investigation of the composition of the 24$\mu$m
counts.  Here we rely on the full optical and {\it Spitzer}-IRAC data
available in the ELAIS N1 field.

\subsection{Optical identifications, photometric redshifts and infrared 
template fits}
\label{sec:templates}

We have associated our band-merged N1 catalogue, which contains all 
 24 $\mu$m detections with a 3.6$\mu$m counterpart to ensure high
 reliability, with the optical UgriZ catalog 
generated from the INT WFS data by
\citet{Babbedge06}, using a search radius of 1.5 arcsec.  This 
search radius is appropriate for SWIRE sources detected at 3.6 $\mu$m 
\citep{mrr05}.   The proportion of 24 $\mu$m blank fields
brighter than S(24) = 200 $\mu$Jy, to the WFS 
survey limit of r $\sim$ 23.5, is 28$\%$. 

All sources with optical associations have been run through 
the photometric redshift code of \cite{mrr03} and \citet{Babbedge04},
with small modifications to the spectral energy distributions (SEDs) 
described by \cite{mrr05}.  20$\%$ of 24 $\mu$m
sources brighter than 200 $\mu$Jy with optical associations failed 
to get a photometric 
redshift because there are less than 4 bands detected at UgriZ, 
3.6, 4.5 $\mu$m or because the reduced $\chi^2$
is too poor ($>$ 10).  This can be because of erroneous optical 
photometry, incorrect optical associations, or because the range of 
optical templates used fails to characterize all sources.  The latter 
possibility is worth further study, especially since, for example, 
strongly reddened quasars or galaxy+quasar  combined SEDs are not included 
in the SED library.

Finally, the SEDs of all 24 $\mu$m sources with photometric redshifts 
are fitted with infrared templates added to the optical/near IR galaxy
and quasar template used in the photometric redshift fit, provided
there is an excess in at least two bands relative to the 
optical/NIR galaxy model.  The templates used are those of 
\citet{mrr04}, also used earlier in \citet{mrr01}: cirrus, 
Arp 220 starburst and a 
mixture of M82 starburst and AGN dust torus.  Examples of SED fits 
using these templates are given by \citet{mrr05}.  
For sources with an infrared excess only at 24$\mu$m, for which 
we can not at this stage characterize the far infrared SED, we 
have fitted an M82 starburst template.  Photometric redshift
distributions of 24$\mu$m sources are shown in \citet{mrr05}.

\subsection{24 $\mu$m counts subdivided by redshift and infrared template type}
\label{sec:subdivided}

\begin{figure*}
\plotone{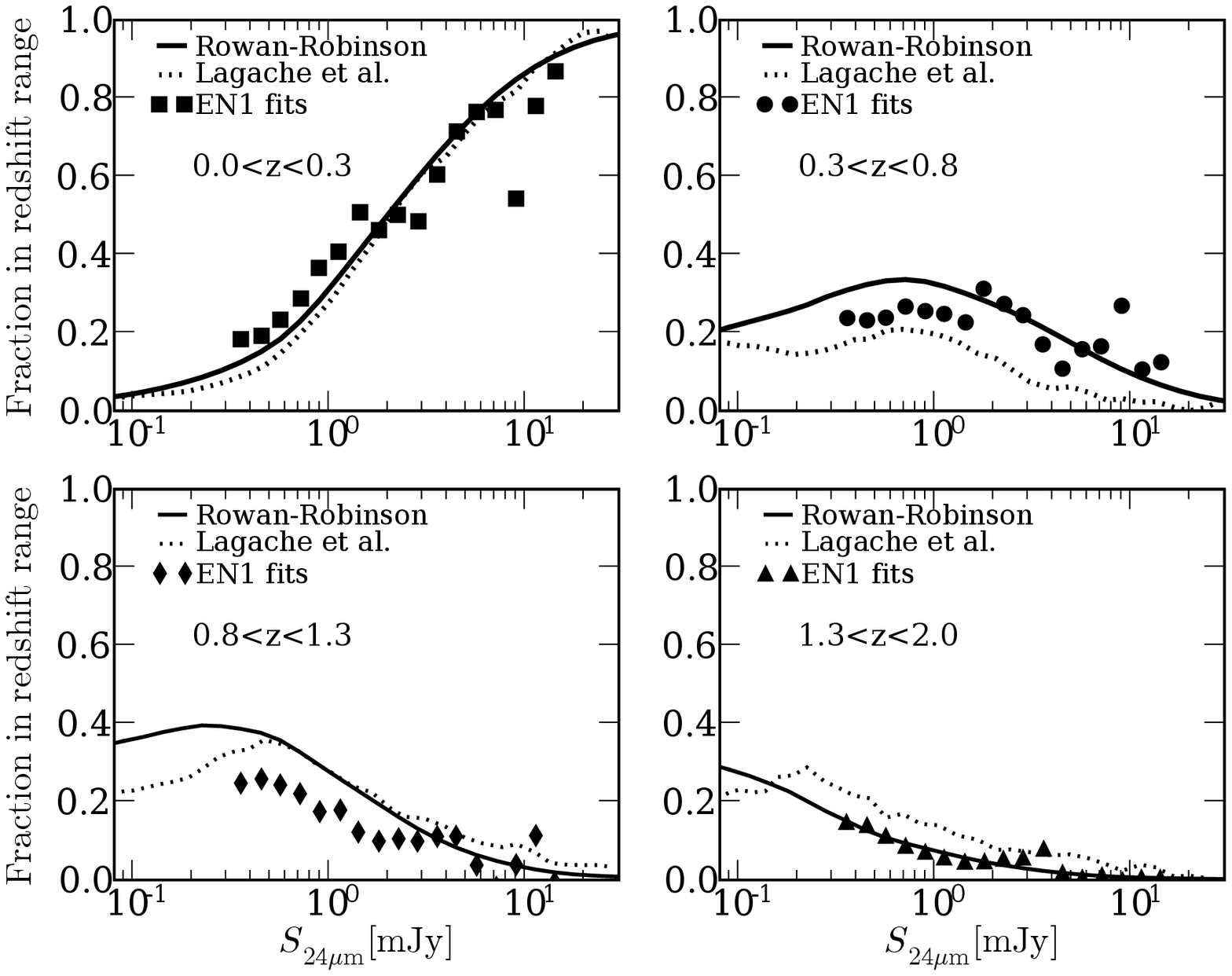}
\caption{Differential counts fraction at 24 $\mu$m subdivided by 
photometric redshift
for the template fits in the SWIRE ELAIS N1 field.  The solid and dashed lines
are derived from the models of \citet{mrr07} and \citet{Lagache04}
respectively.}
\label{fig:zfrac}
\end{figure*}

The results of the template fitting may be used to break down the infrared
counts by redshift range and type.  The fraction of counts per redshift
range is shown in Figure \ref{fig:zfrac}.  
A related plot is shown in
Figure 2 of \citet{Babbedge06} but here we show the results as a fraction of total
counts.  Overlaid are the same fractions taken from the models
of \citet{Lagache04} and \citet{mrr07}.  
The templates fits and the model are generally in
agreement for $z\le 0.3$.  The model yields a higher fraction in the 
higher redshift bins than the template fits; however since some
ELAIS N1 sources did not receive a redshift in the fitting procedure, this
may explain the difference.  Similar results were found by \citet{LeFloch05}
and \citet{Perez05}
in which the models generally underpredicted the sources in the low-redshift
bins and overpredicted the number at higher redshifts.  

\begin{figure*}
\plotone{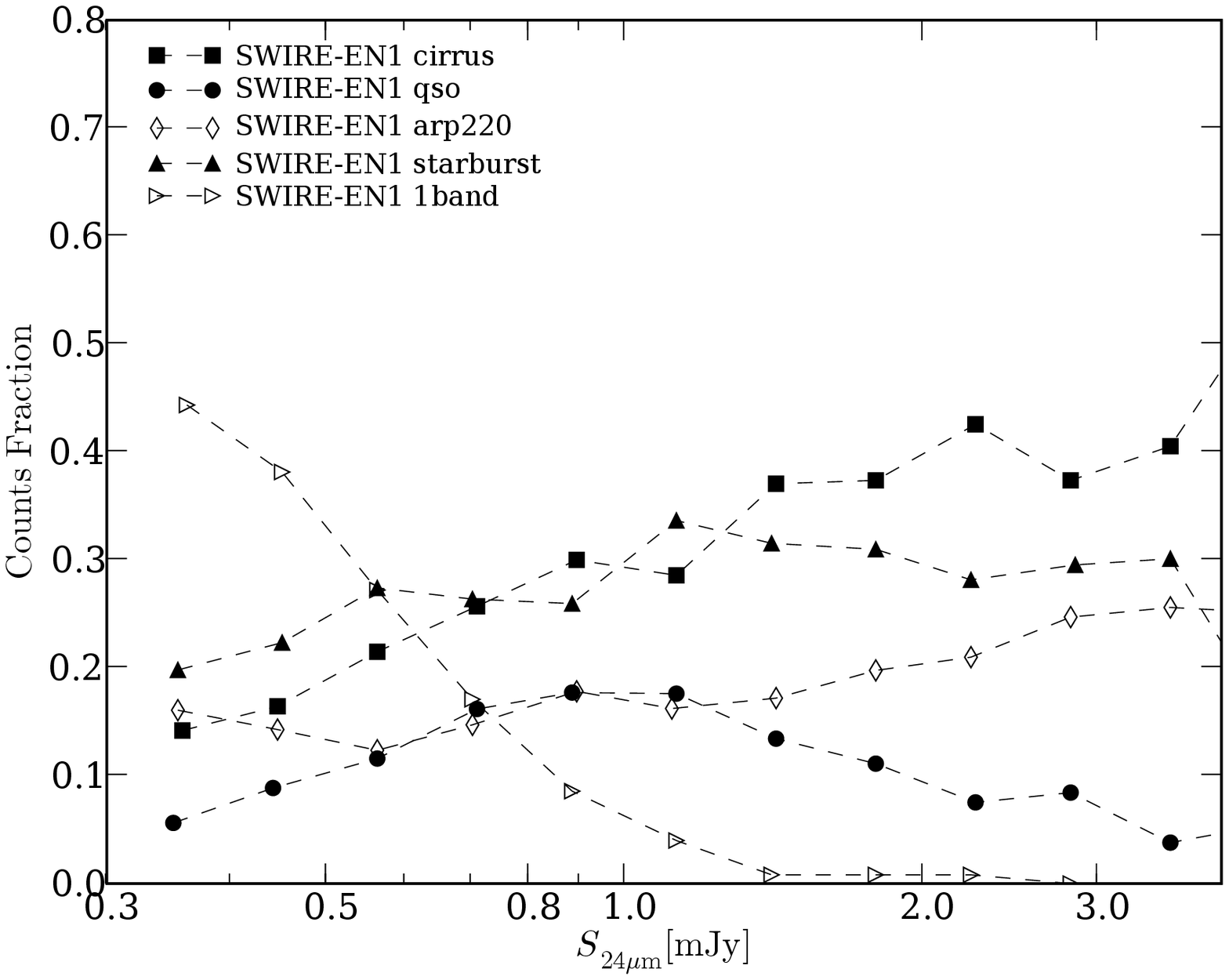}
\caption{Differential counts fraction at 24 $\mu$m subdivided by 
infrared template 
type, for the template fitting of SWIRE ELAIS N1 sources.}  
\label{fig:typefrac}
\end{figure*}

Figure \ref{fig:typefrac} shows the 24 $\mu$m 
differential counts subdivided by 
infrared template type.  We see that the 
steep rise between 1 mJy and 300 $\mu$Jy is caused primarily by 
starbursts, and sources with a infrared excess at 24 $\mu$m only.
To really understand the 24 $\mu$m counts, we need 70 $\mu$m data for
the sources with infrared excess over the fitted template at 24 $\mu$m only.
The fraction of sources with quasar-like infrared SEDs is below about 30\% at
bright fluxes, decreasing to less than 15\% at relatively faint fluxes.

\section{Summary}
\label{sec:summary}

The variation between fields of the SWIRE 24$\mu$m number counts shows 
the importance of sample variance.  For fluxes near a few mJy, the
field-to-field variation in the counts is 50\% to 100\% larger than
is accountable solely from Poisson uncertainties.   Based on redshift estimates from
template fitting and from models, the counts variations near 3 mJy
may result from large-scale structures around $z\sim 0.5$.   At fainter flux
levels, the SWIRE 24$\mu$m catalogs will be useful for tracing structures
at $z\sim 1$ and beyond.

The 24 $\mu$m counts seem exceptionally interesting and most models
published before the launch of {\it Spitzer} miss badly.  There
may be new populations--\cite{mrr05} claim there is a large
population of Type 2 AGN that have not been accounted for in
previous models.  The SWIRE counts confirm a steep rise in
the counts from about 2 to 0.3 mJy that is quite difficult
for models to fit.

Template fitting to the SWIRE EN1 sources is successful for at
least 85\% of the galaxies.  The photometric redshifts indicate
that at least a quarter of the sources in the 1-10 mJy interval
have photometric redshifts in the range 0.3 to 0.8.

The 25$\mu$m band onboard IRAS is closest in wavelength space to
the MIPS 24$\mu$m passband.  IRAS counts are complete to about
250 mJy \citep{Shupe98} and the SWIRE counts go to about 1000x fainter.  
SWIRE results should be  the best 24 $\mu$m counts 
above 250 $\mu$Jy because it will cover the most area of the {\it Spitzer}
extragalactic surveys.  The full area of {\it Spitzer}
surveys solidify the counts up to about 30 mJy, almost allowing 
linking up with the all-sky IRAS counts,
but leaving a gap of a factor of ten in flux.  It
will be left for the {\it Akari (ASTRO-F)} and {\it WISE} missions 
to bridge this gap.

\acknowledgments

We are grateful to the referee for comments, which have led to substantial
improvement of this paper.  This work is based on observations made with 
the Spitzer Space Telescope,
 which is operated by the Jet Propulsion Laboratory, 
California Institute of Technology under a contract with NASA. 
Support for this work was provided by NASA through an award 
issued by JPL/Caltech.
SB was supported by a Fondazione Ing.\ Aldo Gini 2006 grant.


\begin{thebibliography}{}


\bibitem[Babbedge et al.(2004)]{Babbedge04} Babbedge, T. S. R., Rowan-Robinson, M.,
Gonzalez-Solares, E., Polletta, M., Berta, S., P\'erez-Fournon, I., 
Oliver, S., Salaman, D. M., Irwin, M., Weatherley, S. J.\ 2004,
\mnras, 253, 654.

\bibitem[Babbedge et al.(2006)]{Babbedge06} Babbedge, T.S.R.,
Rowan-Robinson, M., Vaccari, M., Surace, J.A., Lonsdale, C.J.,
Clements, D.L., Farrah, D., Fang, F., Franceschini, A., 
Gonzalez-Solares, E., Hatziminaoglou, E., Lacey, C.G., Oliver, S.,
Onyett, N., P\'erez-Fournon, I., Polleta, M., Pozzi, F., 
Rodighiero, G., Shupe, D.L., Siana, B., \& Smith, H.E.\ 2006,
\mnras, 370, 1159.

\bibitem[Berriman et al.(2004)]{montage}Berriman, G.B., Deelman, E., Good, 
J.C., Jacob, J.C., Katz, D.S., Kesselman, C., Laity, A.C.,
Prince, T.A., Singh, G., \& Su, M.\ 2004, ``Ground-based Telescopes,'',
ed.\ Oschmann, J.M. Jr., Proc.\ SPIE, Volume 5493, pp. 221-232.

\bibitem[Berta(2005)]{Berta05} Berta, S.\ 2005, Ph.D.\ Thesis, University
of Padua.

\bibitem[Berta(2006)]{Berta06} Berta, S.\ 2006, \pasp, 118, 754.

\bibitem[Bertin \& Arnout(1996)]{Bertin96} Bertin, E., \& Arnout, S., 1996, 
\aap, 117, 393.

\bibitem[Chary et al.(2004)]{Chary04} Chary, R., Casertano, S., 
Dickinson, M., Ferguson, H., Eisenhardt, P., Elbaz, D., Grogin, N., 
Moustakas, L., Reach, W., Yan, H.\ 2004, \apjs, 154, 80.

\bibitem[Dole, Lagache \& Puget(2003)]{dole03} Dole, H., Lagache, G., 
\& Puget, J.-L., 2003 , \apj, 585, 617.

\bibitem[Elbaz et al.(1999)]{Elbaz99} Elbaz, D. et al.\ 1999, \aap,
351, L37.

\bibitem[Elbaz et al.(2002)]{Elbaz02} Elbaz, D., Cesarsky, C.J.,
Chanial, P., Aussel, H., Franceschini, A., Fadda, D.,
\& Chary, R.R.\ 2002, \aap, 384, 848.

\bibitem[Farrah et al.(2006)]{Farrah06} Farrah, D., Lonsdale, C. J.,
Borys, C., Fang, F., Waddington, I., Oliver, S., Rowan-Robinson, M., 
Babbedge, T., Shupe, D., Polletta, M., Smith, H. E., \& Surace, J.\ 2006,
\apjlett, 641, 117.

\bibitem[Franceschini et al.(2001)]{franceschini01} Franceschini, A. 
et al.\ 2001,  \aap, 378, 1.

\bibitem[Franceschini et al.(1997)]{franceschini97} Franceschini, A.,
Aussel, H., Bressan, A., Cesarsky, C. J., Danese, L., de Zotti, G., Elbaz, D.,
Granato, G. L., Mazzei, P., \& Silva, L.\ 1997,
{\it The Far Infrared and Submillimetre Universe},
ed.\ by A. Wilson, Noordwijk(The Netherlands) ESA SP-401,  p.159.

\bibitem[Franceschini et al.(2007)]{franceschini07} Franceschini, A.,
Berta, S., \& Vaccari, M.\ 2007, in preparation.



\bibitem[Gregorich et al.(1995)]{gregorich95}
Gregorich, D.T., Neugebauer, G.,
Soifer, B.T., Gunn, J.E., \& Herter, T.L. 1995, \aj, 110, 259.

\bibitem[Gruppioni et al.(2005)]{Gruppioni05}
Gruppioni, C., Pozzi, F., Lari, C., Oliver, S., \& Rodighiero, G.\ 2005,
\apjl, 318, 9.

\bibitem[Hacking et al.(1987) Hacking, Houck,\& Condon]{hacking87}
Hacking, P., Houck, J.R., \& Condon, J.J. 1987, \aj, 316, 15.

\bibitem[Hao et al.(2005)]{Hao05} Hao, L., Spoon, H. W. W., Sloan, G. C., 
Marshall, J. A., Armus, L., Tielens, A. G. G. M., Sargent, B., 
van Bemmel, I. M., Charmandaris, V., Weedman, D. W., Houck, J. R.\ 2005,
\apjl, 625, L75.


\bibitem[Houck et al.(2005)]{Houck05} Houck, J.R.,Soifer, B. T., 
Weedman, D., Higdon, S. J. U., Higdon, J. L., Herter, T., Brown, M. J. I.,
Dey, A., Jannuzi, B. T., Le Floc'h, E., Rieke, M., Armus, L., 
Charmandaris, V., Brandl, B. R., Teplitz, H. I.\ 2005, \apjl, 622, L105.


\bibitem[King \& Rowan-Robinson(2002)]{king} King A. J., \& Rowan-Robinson,
M., 2003, \mnras, 339, 260.

\bibitem[Kron(1980)]{Kron80} Kron, R.G.\ 1980, \apjs, 43, 305.

\bibitem[Lagache et al.(2004)]{Lagache04}Lagache, G., 
Dole, H., Puget, J.-L., Perez-Gonzalez, P., Le Floc'h, E., Rieke, G., 
Papovich, C., Egami, E., Alonso-Herrero, A., Engelbracht, C., Gordon, K., 
Misselt, K., \& Morrison, J.\ 2004, ApJS, 154, 112.

\bibitem[Le Floc'h et al.(2005)]{LeFloch05}   Le~Floc'h, E., Papovich, C., Dole, H.,
  Bell, E.~F., Lagache, G., Rieke, G.~H., Egami, E., P{\'e}rez-Gonz{\'a}lez, P.~G., 
  Alonso-Herrero, A., Rieke, M.~J., Blaylock, M., 
Engelbracht, C.~W., Gordon, K.~D., Hines, D.~C., 
Misselt, K.~A., Morrison, \& J.~E., Mould, J.\ 2005, \apj, 632, 169.

\bibitem[Lonsdale \& Chokshi(1993)]{lonsdale93} Lonsdale, C.J., \&
Chokshi, A. 1993, \aj, 105, 1333.

\bibitem[Lonsdale et al.(2003)]{lonsdale03} Lonsdale, C.J., et al 2003, 
\pasp, 115, 897.

\bibitem[Lonsdale et al.(2004)]{lonsdale04} Lonsdale, C.J., et al 2004,
\apjs, 154, 54.


\bibitem[Makovoz \& Marleau(2005)]{makovoz} Makovoz, D., \& Marleau,
F.\ 2005, \pasp, 117, 1113.


\bibitem[Marleau et al.(2004)] {marleau04} Marleau, F., Fadda, D., 
Storrie-Lombardi, L., Helou, G., Makovoz, D., Frayer, D., Yan, L., Appleton,
P., 
Armus, L., Chapman, S., Choi, P., Fang, F., Heinrichsen, I., Im, M., Lacy, M., 
Shupe, D., Soifer, B., Squires, G., Surace, J., Teplitz, H., \& Wilson, G., 
2004, \apjs, 154, 66.

\bibitem[Matute et al.(2006)] {Matute06} Matute, I., La Franca, F., 
Pozzi, F., Gruppioni, C., Lari, C., Zamorani, G.\ 2006, \aap, 451, 443.

\bibitem[MIPS Data Handbook(2006)]{mipshandbook}  MIPS Data Handbook, version 3.2,
06 February 2006, \\
http://ssc.spitzer.caltech.edu/mips/dh/mipsdatahandbook3.2.pdf.

\bibitem[Papovich et al.(2004)]{Papovich04} Papovich, C., Dole, H., Egami, E., 
Le Floc'h, E., Perez-Gonzalez, P., Alonso-Herrero, A., Bai, L., Beichman, C., 
Blaylock, M., Engelbracht, C., Gordon, K., Hines, D., Misselt, K., Morrison, J., 
Mould, J., Muzerolle, J., Neugebauer, G., Richards, P., Rieke, G., Rieke, M., 
Rigby, J., Su, K., \& Young, E., 2004, \apjs, 154, 70.

\bibitem[P\'erez-Gonz\'alez et al.(2005)]{Perez05}  P{\'e}rez-Gonz{\'a}lez, P.~G.,
 Rieke, G.~H., Egami, E., 
Alonso-Herrero, A., Dole, H., Papovich, C., Blaylock, M., 
Jones, J., Rieke, M., Rigby, J., Barmby, P., 
Fazio, G.~G., Huang, J., \& Martin, C.\ 2005, \apj, 630, 82.
 
\bibitem[Polletta et al.(2006)]{Polletta07} Polletta, M. et al.\ 2007,
in preparation.

\bibitem[Pozzi et al.(2004)]{Pozzi04} Pozzi, F., Gruppioni, C., Oliver,
S., Matute, I., La Franca, F., Lari, C., Zamorani, G., Franceschini,
A., \& Rowan-Robinson, M., 2004, \apj, 609, 122. 

\bibitem[Rieke et al.(2004)]{rieke04} Rieke, G. et al.\ 2004,
\apjs, 154, 25.


\bibitem[Rodighiero et al.(2006)] {Rodighiero06} Rodighiero, G., Lari, C., Pozzi, F., Gruppioni, C.,
Fadda, D., Franceschini, A., Lonsdale, C., Surace, J., Shupe, D., \& Fang, F.\ 2006,
\mnras, 371, 1891.

\bibitem[Rowan-Robinson(2001)]{mrr01} Rowan-Robinson, M.\ 2001, \apj,
549, 745.

\bibitem[Rowan-Robinson(2003)]{mrr03} Rowan-Robinson, M.\ 2003, \mnras, 345, 819.

\bibitem[Rowan-Robinson et al.(2004)]{mrr04} Rowan-Robinson, M., et al.\ 2004,
\mnras, 351, 1290.

\bibitem[Rowan-Robinson et al.(2005)]{mrr05} Rowan-Robinson, M. et al.\ 2005, \aj, 129, 1183 .

\bibitem[Rowan-Robinson(2007)]{mrr07} Rowan-Robinson, M.\ 2007, 
in preparation.

\bibitem[Skrutskie et al.(2006)]{Skrutskie06} 
Skrutskie, M.F., Cutri, R.M., Stiening, R., Weinberg, M.D., Schneider, S.,
Carpenter, J.M., Beichman, C., Capps, R., Chester, T., Elias, J.,
Huchra, J.,  Liebert, J., Lonsdale, C., Monet, D.G., Price, S., Seitzer, P.,
Jarrett, T., Kirkpatrick, J.D., Gizis, J., Howard, E., Evans, T.,
Fowler, J., Fullmer, L.,  Hurt, R., Light, R., Kopan, E.L.,
Marsh, K.A., McCallon, H.L., Tam, R., Van Dyk, S., \&  Wheelock, S.\ 2006, 
\aj, 131, 1163.

\bibitem[Shupe et al.(1998)]{Shupe98} Shupe, D.L., Fang, F.,
Hacking, P.B., \& Huchra, J.P.\ 1998, 501, 597.

\bibitem[Surace et al.(2007)]{Surace06}
Surace, J.A., et al. 2007, in preparation.

\bibitem[Vaccari et al.(2005)]{Vaccari05} Vaccari, M., Lari, C., 
Angeretti, L., Fadda, D., Gruppioni, C., Pozzi, F., Prouton, O., 
Aussel, H., Babbedge, T., Ciliegi, P., Franceschini, A., 
Gonzalez-Solares, E., La Franca, F., Oliver, S., P\'erez-Fournon, I., 
Rowan-Robinson, M., Serjeant, S., V\"ais\"anen, P.\ 2005,
\mnras, 358, 397.

\bibitem[Xu et al.(1998)]{Xu98} Xu, C., Hacking, P.B., Fang, F.,
Shupe, D.L., Lonsdale,C.J., Lu, N.Y., Helou, G., Stacey, G.J., \&
Ashby, M.L.N.\ 1998, \apj, 508, 576.


\bibitem[Xu et al.(2003)]{Xu03} Xu, C.K., Lonsdale, C.J.,
Shupe, D.L., Franceschini, A., Martin, C., \& Schiminovich,
D. 2003, \apj, 587, 90.

\bibitem[Xu et al.(2007)]{Xu06} Xu, C.K., et al.\ 2007, in
preparation.


\end{thebibliography}
\end{document}